\documentclass{article}[15pt]

\usepackage[dvips]{epsfig}
\usepackage{rotating}

\begin{document}

\title{
How Many Genes Are Needed for a  Discriminant \\
Microarray Data Analysis ? \\
\vspace{0.2in}
\author{
Wentian Li and  Yaning Yang \\
{\small \sl  Laboratory of Statistical Genetics, Box 192} \\
{\small \sl The Rockefeller University, 1230 York Avenue, New York, NY 10021, USA}
} 
\date{March 12, 2001}
} 
\maketitle    
\markboth{\sl Li,Yang}{\sl Li Yang}

\vspace{-0.2in}

\begin{abstract}
The analysis of the leukemia data from Whitehead/MIT group is a 
discriminant analysis (also called a supervised learning). Among 
thousands of genes whose expression levels are measured, not all 
are needed for discriminant analysis: a gene may either not contribute 
to the separation of two types of tissues/cancers, or it may be 
redundant because it is highly correlated with other genes. There 
are two theoretical frameworks in which variable selection (or gene 
selection in our case) can be addressed. The first is model selection, 
and the second is model averaging. We have carried out model 
selection using Akaike information criterion and Bayesian information 
criterion with logistic regression (discrimination, prediction, or 
classification) to determine the number of genes that provide the 
best model. These model selection criteria set upper limits of 
22-25 and 12-13 genes for this data set with 38 samples, and 
the best model consists of only one (no.4847, zyxin) or two genes. 
We have also carried out model averaging over the best single-gene 
logistic predictors using three different weights: maximized likelihood, 
prediction rate on training set, and equal weight. We have observed 
that the performance of most of these weighted predictors on the 
testing set is gradually reduced as more genes are included, but a clear 
cutoff that separates good and bad prediction performance is not found.
\end{abstract}

\newpage


\section*{Introduction}

\indent

There are two types of microarray experiments. In the first type, 
samples are not labeled, but gene expression levels are followed 
with time. The goal is to find genes whose expression levels move 
together. In the second type, samples are labeled as either normal 
or diseased tissues. The goal is to find genes whose expression levels 
can distinguish different labels. Using terminology from machine learning, 
the data analysis of the first type of experiment is ``unsupervised", 
whereas that of the second type is ``supervised" (see, e.g., 
[Haghighi, Banerjee, Li, 1999]). The prototype of the first type 
of analysis is cluster analysis, whereas that of the second type 
is discriminant analysis.

We focus on the data analysis of the second type, using the leukemia 
data from Whitehead/MIT's group [Golub et al, 1999]. The question 
we are addressing is how many gene expressions are needed for a 
discriminant analysis. In this data set, 7129 gene expressions are 
measured on 38 sample points. In the original analysis [Golub et al, 
1999], 50 out of more than 7000 gene expressions are used for 
discriminant analysis. We ask whether 50 genes are still too 
large a number, considering that the sample size is only 38. 
This seemingly simple question may receive two answers from two 
different perspectives: model selection (see, e.g., 
[Burnham, Anderson, 1998]) and model averaging (see, e.g., 
[Geisser, 1974]). We will address the two separately below.

\section*{Model Selection}

\indent

The goal of model selection is to pick the model among many possible 
models that achieves the best balance between data-fitting and model 
complexity. A perfect data-fitting performance by a complicated model 
with many parameters can well be an example of overfitting. On the 
other hand, a simplistic model with few parameters that fits the 
data poorly is an example of underfitting. There are two proposals 
for achieving the balance between data-fitting and model complexity: 
Akaike information criterion (AIC) [Akaike, 1974; Parzen, Tanabe, 
Kitagawa, 1998; Burnham, Anderson, 1998] and Bayesian information 
criterion (BIC) [Schwarz, 1976; Raftery, 1995]. These two quantities 
are defined as:
\begin{equation}
\label{aic-bic}
\mbox{AIC} = -2 \log (\hat{L}) + 2K, 
\hspace{0.2in} 
\mbox{ and}
\hspace{0.2in} 
\mbox{BIC} = -2 \log (\hat{L}) + \log(N) K ,
\end{equation}
where $\hat{L}$ is the maximized likelihood, $K$ the number
of free parameters in the model, and $N$ the sample size. High-order
terms of $O(1/N)$ (for AIC) and $O(1)+ O(1/\sqrt{N}) +O(1/N)$ (for BIC)
are ignored here for simplicity. The model with the lowest AIC or 
BIC is considered to be the preferred model. 

Selecting genes relevant to a discriminant microarray data analysis may 
become an issue of model selection. But we have to be clear in what 
context this is the case. Model selection adjusts the number of 
parameters in the model, not the number of variables. Nevertheless, 
if we plan to combine variables (additively or any other functional 
form), each variable has one or perhaps more coefficient. Removing 
a variable removes the corresponding coefficient. In this context, 
variable selection is a special case of model selection. We illustrate 
this by the logistic regression/discrimination for our data set:
\begin{equation}
\label{lr}
\mbox{Prob(AML)} = 
\frac{1}{1+ e^{ -a_0 - \sum_{j \in top \hspace{0.05in} genes} a_j x_j }}
\end{equation}
where $x_j$ is the (log, normalized) gene expression level of gene $j$ 
among the top performing genes, and AML (acute myeloid leukemia)
is one type of leukemia (ALL, the acute lymphoblastic leukemia, is
another type). Using a linear combination of all 7129  genes in 
logistic discrimination requires 7130 parameters, whereas using one 
gene requires only 2 parameters. AIC/BIC defined in Eq.(\ref{aic-bic}) 
will then compare multiple-gene models with single-gene models, 
and determine which scenario is better.

Our results are summarized in Table 1, where we list the type of 
model (which variables are additively combined in the logistic regression), 
number of parameters in the model, -2log of the maximum likelihood, 
AIC/BIC (absolute and relative), prediction rate in training set, 
and that in the testing set. The most striking result from this data 
set is that many genes are strong predictors for the leukemia class, 
consistent with the observation in [Golub et al, 1999]. This observation 
is confirmed by the following facts in Table 1: logistic regressions 
using the top 2, 5, 10, 22,and 37 genes all fit the data almost perfectly 
(as measured by the $-2\log(\hat{L})$ value) (note that the model is saturated 
when the number of variables used in the logistic regression is 37, 
since the number of parameters is then equal to the number of sample 
points); stepwise variable selection leads to only two genes (note 
that stepwise variable selection fails to find the best model, a 
single-gene model, because it is a local minimization/maximization 
procedure); the No.10 best-performing gene is only slightly worse 
than the No.1 performing gene: 35 vs. 36 correct predictions on the 
training set (though the No.100 and No.200 best-performing genes predict 
only 30 and 29 correctly on the training set); etc. This situation 
of strong prediction or easy classification is in contrast with the 
epidemiology and pedigree data used in human complex diseases, where 
strong predictors are rather rare [Li, Sherriff, Liu, 2000; Li, Nyholt, 2001].

Thest two rows in Table 1 are predictors/classifiers that do not use 
any gene expression levels. These are random guesses or null models. 
The first null model (\#1) uses the proportion of AML in all samples 
as Prob(AML) (11/38 in our training set) as the guessing probability 
for AML. The second null model (\#0) uses half-half probabilities. 
It is interesting to note that to beat both null, the number of genes 
in logistic regression can not exceed some upper limits: for null 
model 1, since we require AIC/BIC to be smaller (where $p$ is the 
number of genes used in logistic regression, and  $-2\log(\hat{L})$
is assumed to be zero for the best-case scenario):
\begin{eqnarray}
0 + 2(p+1) \approx  \mbox{AIC} & <  &\mbox{AIC (fixed)} = 47.728
\nonumber \\
0 + 3.637568 (p+1) \approx  \mbox{BIC} & < & \mbox{BIC (fixed)} = 49.365,
\end{eqnarray}
it sets $p < 22.86 \approx 22 $, and $p < 12.57 \approx 12 $,
respectively. Similarly, to beat the random-guess model, we require:
\begin{eqnarray}
0 + 2(p+1) \approx  \mbox{AIC} & <  &\mbox{AIC (random)} = 52.679
\nonumber \\
0 + 3.637568 (p+1) \approx  \mbox{BIC} & < & \mbox{BIC (random)} = 52.679
\end{eqnarray}
which are $ p < 25.34 \approx 25 $, and $p < 13.48 \approx 13$, respectively. 
These specific upper limits are directly related to the fact that the 
sample size is 38. They will move up and down with the sample size.

\begin{table}
\begin{center}

\begin{tabular}{|c|c|c|c|c|c|c|c|c|}
\hline
type & K & -2log($\hat{L}$) & AIC & $\Delta$AIC &  BIC & $\Delta$BIC
 & $p_{train}$ & $p_{test}$ \\
\hline
\#1 g4847 (zyxin) &  2 & 5$\times$10$^{-9}$$^{(a)}$& 4.000& 0 & 7.275 & 0 &
	38/38 & 31/34  \\
\#2 g1882 (CST3 cystatin C) &  2 & 6.973& 10.973 & 6.973 & 14.248& 6.973&
	36/38 & 32/34 \\
\#3 g3320 (leukotriene c4 synthase)&2& 10.914& 14.914& 10.914& 18.190&10.915& 
	35/38 & 27/34 \\ 
\#4 g5039 (LEPR leptin receptor) & 2& 11.355& 15.355& 11.355& 18.630 & 11.355& 
	36/38 & 22/34 \\ 
\#5 g6218 (ELA2 elastatse 2) &2& 11.459& 15.459& 11.459& 18.734& 11.459& 
	34/38 & 22/34 \\ 
\#6 g2020 (FAH ..)& 2& 12.103& 16.103& 12.103& 19.378&12.103& 
	36/38 & 25/34 \\ 
\#7 g1834 (CD33 antigen) &  2 & 12.226 & 16.226& 12.226& 19.501&12.226& 
	35/38 & 31/34 \\ 
\#8 g760 (cystatin A)  &  2 & 13.104 & 17.104&  13.104 & 20.379 &13.104& 
	35/38 & 32/34 \\ 
\#9 g1745 (LYN v-yes-1..) &  2 & 13.151 & 17.151& 13.151& 20.426&13.151&
	33/38 & 28/34 \\ 
\#10 g5772 (c-myb) &  2 & 14.723 & 18.723  & 14.723& 21.998 & 14.723& 
	35/38 & 27/34 \\ 
\#100 g2833(AF1q) &  2 & 27.215 &  31.215 & 27.215& 34.490 & 27.215& 
	30/38 & 28/34 \\ 
\#200 g3312(protein kinase ATR) &  2& 30.841& 34.841& 30.841& 38.117 &30.842& 
	29/38 & 21/34 \\ 
\hline
g1834+g2267$^{(b)}$ & 3 & 0.004 & 6.004 
	&2.004 & 10.917 & 3.642& 38/38  & 22/34  \\ 
g5039+g5772$^{(c)}$ & 3 & 0.008 & 6.008
	& 2.008 & 10.921 & 3.646& 38/38 & 26/34 \\
\hline
top 2 (g4847+g1882) & 3 & 0.029 & 6.029& 2.029& 10.942 &3.667& 
	38/38 & 32/34\\
top 5 & 6 & 0.011 & 12.011& 8.011 & 21.837  & 14.562& 38/38 & 24/34\\
top 10 & 11 & 0.002 & 22.002 & 18.002  & 40.016 & 32.741& 
	38/38 & 31/34\\
top 22 & 23 & 0.001 & 46.001 & 42.001  &  83.666 & 76.391& 
	38/38 & 27/34\\
top 37 & 38 & 0.001 & 76.001 & 72.001 & 138.229 & 130.954& 
	38/38 & 21/34\\
\hline
null 1 (proportion estimation)&  1 & 45.728$^{(d)}$ & 47.728 & 43.728& 49.365 & 42.090& 
	27/38$^{(d)}$ & 20/34$^{(d)}$ \\
\hline
null 0 (random guess) & 0  & 52.679$^{(e)}$ & 52.679 & 48.679 & 52.679 & 45.404  & 
	19/38$^{(e)}$  & 17/34$^{(e)}$ \\
\hline
\end{tabular}

\end{center}
\caption{
Logistic regression results (sample size N=38, log(N)=3.637568).
$K$: number of free parameters in the logistic regression (LR) 
(number of genes included plus 1); $\Delta$AIC ($\Delta$BIC) is 
the AIC (BIC) value relative to that of the best model
(single-gene LR using g4847, zyxin); $p_{train}$ ($p_{test}$)
is the prediction rate on the training (testing) set; 
``top 37" is the LR using 37 best genes by their single-variable 
LR performance. Notes 
(a) Since this model/predictor fits the data perfectly, $\hat{L}$
should be 1, and $-2\log(\hat{L})$  should be 0. In a real optimization 
procedure, the actual value may depend on the number of iterative 
steps. 
(b) This LR is selected by a stepwise variable selection (by either 
AIC or BIC) from the starting group of top 22 genes (top 22
$\rightarrow$ A/BIC $\rightarrow$ 2);
(c) Similar to (b), but the starting group of genes in the LR 
containing the top 10 genes (top 10 $\rightarrow$ A/BIC $\rightarrow$ 2);
(d) This null model uses 11/38 $\approx$  0.29 as the probability for 
AML for any sample data. Since 0.29 $<$ 0.5, any sample will be 
predicted as ALL type, which is correct 27 times in the training set, 
and 20 times in the testing set. The $-2\log(\hat{L})$ is
equal to  $-2 \log [ (27/38)^{27} (11/38)^{11} ]$;
(e) The   $-2\log(\hat{L})$, $p_{train}$, and $p_{test}$ 
for the random guess model is expected to be:
$-2\log(0.5^N) = 2N\log(2)$, 0.5, and 0.5.
}
\end{table}

\section*{Model Averaging}

\indent

In the model selection framework, we can not use a logistic regression 
with too many genes because it may not improve the data-fitting performance 
enough to compensate for the increase of model complexity. This conclusion 
is correct when one model (e.g. a logistic regression that uses one 
gene) is compared to other alternative models (e.g. a logistic regression 
that uses, say, 10 genes). Nevertheless, it is possible to average/combine 
many different models each involving one gene. The restriction on model 
complexity during the model selection process does not apply to model 
averaging. Model averaging has also been discussed under names such as 
``committee machines", ``boosting weak predictors", ``mixture of experts", 
``stacked generalization" (see, e.g., [Ripley, 1996]).

Without guidance from the model selection framework on the number 
of models (number of genes) to be included, we have tried several 
empirical approaches. We first examine whether there is a gap in 
data-fitting performance among top genes. Genes that do not fit the 
data should not be considered in model averaging. For this purpose, 
Fig.1 shows the  $-2 \log(\hat{L})$   for the top 1000 genes on the 
training set, as a function of the (log) rank. A linear fitting of 
$-2\log(\hat{L})$ on $\log$(rank)  seems to fit the points well, and 
it is hard to see ``better-than-average" genes except the first gene. 
This is not surprising since g4847 discriminates the 38 training sample 
points perfectly. Note that the linear trend in Fig.1 is equivalent to 
a power-law function in likelihood vs. rank plot ( $\hat{L} \sim 1/r^{2.56}$).
Such power-law is similar to the power-law rank-frequency plots observed 
in many social and natural data, also known as Zipf's law 
[Zipf, 1949; Li, 1997-2001]. In any case, there is no discernible gap 
in Fig.1 that separates relevant and irrelevant genes.

We then check how the prediction rate on the training set correlates 
with that on the testing set. Fig.2 shows the error rates on both training 
(x-axis) and testing sets (y-axis) for the top 500 performing genes. 
The left plot in Fig.2 shows the mean square error, and the right plot 
shows the prediction errors. The left plot contains both information on 
the success rate of prediction and that of confidence of prediction, 
whereas the right plot contains only information on the success rate 
of prediction. Points along the diagonal line in Fig.2 exhibit similar 
error rates in the training and testing set, and thus are reasonable 
predictors. On the other hand, points well above the diagonal line 
indicate overtraining. The most reasonable predictor based on both the 
training and testing set is g1882 (on the other hand, the best predictor 
based on training set is g4847).

Finally we examine the model averaging performance with various numbers 
of models included, each being a single-gene logistic regression:
\begin{equation}
\mbox{Prob(AML)} =
\label{model-ave}
\sum_{j \in \mbox{top genes}} w_j \left( \frac{1}{1+e^{-a_j-b_j x_j}} \right).
\end{equation}
We have chosen three weighting schemes: the first is proportional to 
the prediction rate on the training set ($w_j \propto p_{j,train}$),
the second is the equal weight ($w_j \propto 1 $),  and the last one 
is proportional to the maximum likelihood as obtained from the training set 
 ($ w_j \propto \hat{L}_j$).  Fig.3-4 shows the behavior of all three 
weighting schemes on both the training and the testing sets, using 
either the mean square error or the prediction error, up to 200 genes. 

The maximum likelihood weight is equivalent to the Akaike weight 
($w_j \propto exp(- \mbox{AIC}_j/2)$)  [Parzen, Tanabe, Kitagawa, 1998] 
and Bayesian weight  ($w_j \propto exp(- \mbox{BIC}_j/2)$) 
[Raftery, 1995], since all models being averaged in Eq.(\ref{model-ave}) 
have the same number of parameters and same sample size (assuming high-order 
terms are ignored). Using Bayesian weight in model averaging is essentially 
a derivation of the posterior predictive distribution [Gelman, et al, 1995]: 
``posterior" because data in a training set is used, and ``predictive"
because unknown new data in the testing set is to be predicted.

For our data set, this weighting scheme is nevertheless uninteresting: the 
mean square error only decreases slightly with the number of models used, 
while the error in prediction rate is unchanged. The reason for this is 
very simple: the best model (the single-gene logistic regression using 
the g4847) discriminates the training set perfectly. Its likelihood is 
much higher than any other models. As a result, the weight of other models 
is negligible, and the model averaging essentially remains as one model. 
The equal weight can be potentially incorrect since models that do not 
fit the data should not contribute as equally as models that fit the data 
better. In our data set, however, the equal weight scheme is actually similar 
to the weighting scheme that uses the prediction rate, because many genes 
(up to 200) exhibit similar prediction rates on the training set.

It is difficult from Fig.3-4 to determine a cutoff on the number of 
models (number of genes) to be included. Fig.3-4, however, clearly 
shows that a lower number of models (genes) is typically better (except 
the maximum likelihood weight, which is insensitive to the number 
of genes included). Fig.3, in particular, indicates that it is 
possible to perform better (in term of mean square error) on the 
testing set than using just one model if the number of models is 
less than around 25. Of course, this result is obtained from the 
specific testing set we have at hand. A more conclusive result may 
require a combining of the training and testing set, or requires more 
sample points than are currently available.

Due to the weight in a model averaging, the apparent number of terms 
(models, genes) included does not reflect on the true number of genes 
involved. For this reason, we may introduce a quantity called ``effective 
number of genes (terms, models)". The leading term in a model averaging 
contributes a number of 1 to this quantity, but the contribution from 
the term j is equal to $w_j/w_1$. For example, with the maximum likelihood 
weight, the relative weights of the first ten terms are 1, 0.031, 
0.0043, 0.0034, 0.0032, 0.0024, 0.0022, 0.0014, 0.0014, and 0.0006. 
The effective number of terms when the top ten genes are included is 
1.05, a far less number than 10.

The discrimination used in [Golub et al, 1999] is a model averaging 
instead of a model selection. The number of genes used is 50. To quote 
from [Golub, et al, 1999], ``the number was well within the total number 
of genes strongly correlated with the class distinction, seemed likely 
to be large enough to be robust against noise, and was small enough to 
be readily applied in a clinical setting." Our Fig.3-4 shows that 
although the prediction rate on the training set stays close to 100\% 
even as the number of models (genes) is increased, the prediction rate 
on the testing set decreases with more genes. The only exception is 
the maximum likelihood weight, where the prediction rate is almost 
unchanged due to the dominance of the best gene. It seems that we should 
not increase the number of models (genes) in model averaging arbitrarily. 

In conclusion, due to the small sample size and the presence of strong 
predictors, we believe the number of genes used in a discriminant analysis 
in this data set can be much smaller than 50. Although we can not give 
a definite answer as to the exact number of genes to be used, one 
proposal is to use only one or two genes, and other exploratory data 
analyses indicate an upper limit of 10-20 genes. Similar analysis of 
other data sets for cancer classifications using microarray will be 
discussed in [Li, et al. 2001] and Zipf's law in these data sets will 
be discussed in [Li, 2001].

\section*{Acknowledgements}

\indent

W.Li's work was supported by NIH grant K01HG00024 and Y. Yang's 
work was supported by the grant MH44292.

\newpage

\section*{References}


\vspace{0.07in}
\noindent
H Akaike (1974), ``A new look at the statistical model identification", 
IEEE Transactions on Automatic Control, 19:716-723.

\vspace{0.07in}
\noindent
KP Burnham, DR Anderson (1998), 
{\sl Model Selection and Inference} (Springer).

\vspace{0.07in}
\noindent
A Gelman, JB Carlin, HS Stern, DB Rubin (1995), 
{\sl Bayesian Data Analysis} (Chapman \& Hall).

\vspace{0.07in}
\noindent
S Geisser (1993), 
{\sl Predictive Inference: An Introduction} (Chapman \& Hall).

\vspace{0.07in}
\noindent
TR Golub, DK Slonim, P Tamayo, C Huard, M Gaasenbeek, JP Mesirov, 
H Coller, ML Loh, JR Downing, MA Caligiuri, CD Bloomfield, ES Lander
(1999),
``Molecular classification of cancer: class discovery and class
prediction by gene expression monitoring",
Science, 286:531-537.

\vspace{0.07in}
\noindent
F Haghighi, P Banerjee, W  Li (1999),
``Application of artificial neural networks in whole-genome
analysis of complex diseases" (meeting abstract), 
Cold Spring Harbor Meeting on Genome Sequencing \& Biology, 
page 75.

\vspace{0.07in}
\noindent
W Li (1997-2001), An online resource on Zipf's law (URL:
http://linkage.rockefeller.edu/wli/zipf/).

\vspace{0.07in}
\noindent
W Li (2001), ``Zipf's  law in importance of genes for cancer 
classification using microarray data",  submitted.

\vspace{0.07in}
\noindent
W Li, A Sherriff, X Liu (2000),
``Assessing risk factors of complex diseases by Akaike
information criterion and Bayesian information criterion" 
(meeting abstract), American Journal of Human Genetics,
67 (supp 2), page 222.

\vspace{0.07in}
\noindent
W Li, D Nyholt (2001),
``Marker selection by AIC/BIC",
Genetic Epidemiology, in press.

\vspace{0.07in}
\noindent
E Parzen, K Tanabe, G Kitagawa (1998), 
{\sl Selected Papers of Hirotugu Akaike} (Springer).

\vspace{0.07in}
\noindent
AE Raftery (1995), ``Bayesian model selection in social research",
in {\sl Sociological Methodology}, ed. PV Marsden (Blackwells),
pages 185-195.

\vspace{0.07in}
\noindent
BD Ripley (1996),
{\sl Pattern Recognition  and Neural Networks}
(Cambridge University Press).

\vspace{0.07in}
\noindent
G Schwarz (1976), ``Estimating the dimension of a model",
Annals of Statistics, 6:461-464.

\vspace{0.07in}
\noindent
GF Zipf (1949), 
{\sl Human Behavior and the Principle of Least Effect}
(Addison-Wesley).

\newpage

\begin{figure}
\begin{center}
  \begin{turn}{-90}
  \epsfig{file=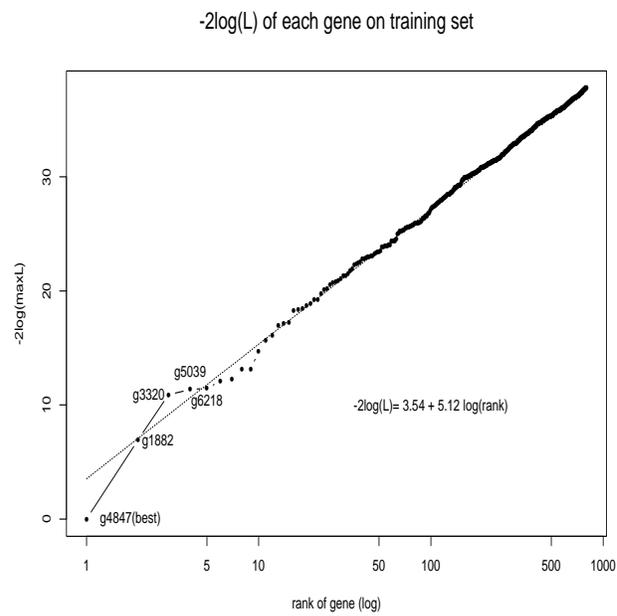, height=8cm, width=8cm}
  \end{turn}
\end{center}
\caption{
Data-fitting performance as measured by $-2\log (\hat{L})$ (the smaller,
the better fit) for the top 1000 genes. The x-axis is the rank of
the gene by the likelihood of its single-gene logistic regression. 
A linear regression line of $-2\log(\hat{L})$ vs. $\log$(rank) is 
also shown, with the slope equal to 5.12.
}
\label{fig1}
\end{figure}

\begin{figure}
\begin{center}
  \begin{turn}{-90}
  \epsfig{file=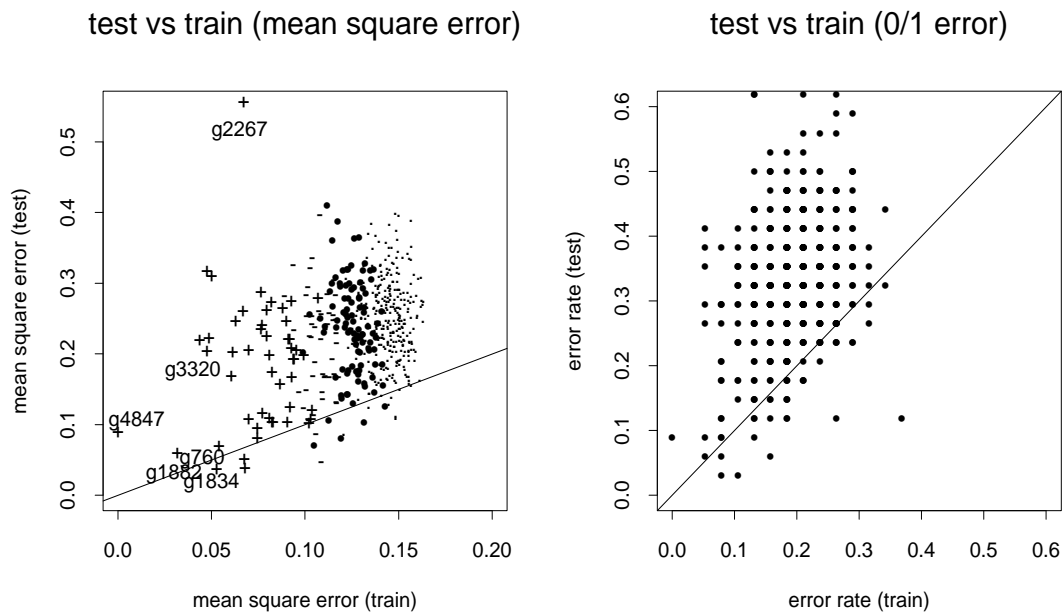, height=14cm, width=8cm}
  \end{turn}
\end{center}
\caption{
Error rates on the testing set (y-axis) are compared with those
on the training set (x-axis) for the top 500 genes (based on the
performance on the training set).  Each point represents 
one single-gene logistic regression predictor. The left plot uses
the mean squared error (``+" for the first 50 top genes, ``-" for
the top 51-100 genes, ``*" for the top 101-200 genes, and ``." for the
top 201-500 genes), and the right plot uses the prediction error.
}
\label{fig2}
\end{figure}

\begin{figure}
\begin{center}
  \begin{turn}{-90}
  \epsfig{file=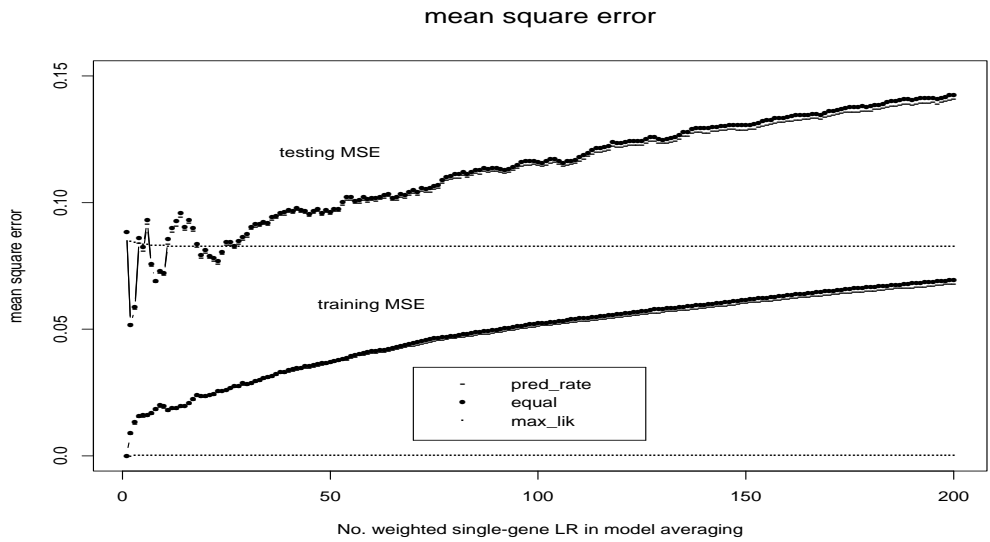, height=13cm, width=7cm}
  \end{turn}
\end{center}
\caption{
Model averaging performance with three different weighting schemes:
(1) weight being proportional to the prediction rate on the training set; 
(2) equal weight; and (3) weight being proportional to the maximum likelihood
obtained on the training set. The mean square error is plotted against 
the number of models being averaged. Each model is a single-gene logistic 
regression.  The mean square error on both the training set (bottom) and 
the testing set (top) are shown.
}
\label{fig3}
\end{figure}

\begin{figure}
\begin{center}
  \begin{turn}{-90}
  \epsfig{file=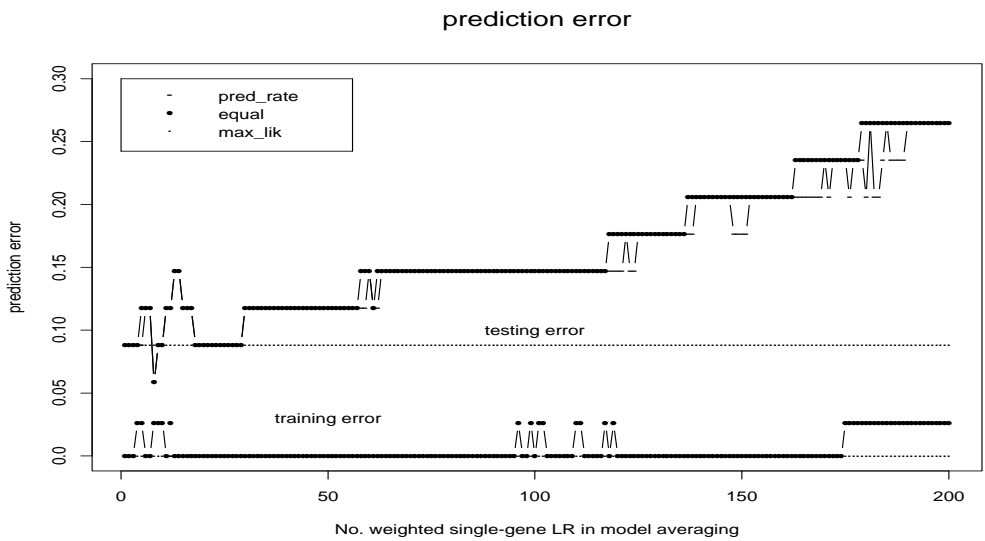, height=13cm, width=7cm}
  \end{turn}
\end{center}
\caption{
Similar to Fig.3, but the prediction  error rate is plotted
against the number of models being averaged (each model is a
single-gene logistic regression).
}
\label{fig4}
\end{figure}

\end{document}